\documentclass[preprint,12pt]{elsarticle}

\usepackage[T2A]{fontenc}
\usepackage[utf8x]{inputenc}

\usepackage{amssymb}
\usepackage{amsmath}
\usepackage{bm}

\journal{{\rm arxiv.org}}

\begin{document}

\begin{frontmatter}



\title{Evolutionarity of MHD shock waves in collisionless plasma with heat fluxes}


\author[]{V.D. Kuznetsov}
\author[]{A. I. Osin}
\ead{osin@izmiran.ru}
 
\affiliation{organization={IZMIRAN},
            addressline={Kaluzhskoye hwy. 4}, 
            city={Troitsk},
            postcode={108840}, 
            state={Moscow},
            country={Russia}}

\begin{abstract}
The evolutionarity conditions for the MHD shock waves are considered within the framework of the 8-moment approximation for collisionless plasma with heat fluxes. In the general case, evolutionarity diagrams are obtained depending on the relative magnitude of the Alfven wave velocity in front of or behind the shock wave front.
The evolutionarity conditions for parallel shock waves are analyzed using previously obtained solutions for parallel MHD shock waves in collisionless plasma with heat fluxes. On the plane of dimensionless parameters characterizing plasma velocity and heat flux in front of the shock wave, the regions of evolutionarity are determined for the fast and slow shock waves propagating along the magnetic field.
\end{abstract}



\begin{keyword}
collisionless plasma \sep shock waves\sep heat fluxes \sep evolutionarity


\end{keyword}

\end{frontmatter}


\section{Introduction}
\label{}

Observations of solar wind plasma discovered non-maxwellian particle distribution and heat fluxes \cite{Hundhausen1968,Whang1971} which should be taken into account in the study of linear wave 
phenomena as well as shock waves. The account of heat fluxes along the magnetic field leads to the 
system of 8-moment approximation for the collisionless plasma \cite{Oraevskii1968,Namikawa1981,OraKonKhaz1985,Zakharov1988tl,Ramos2003,Kuznetsov2009en}.
Small amplitude waves in this approximation have been studied in 
\cite{Namikawa1981,Barkhudarov1987tl,Zakharov1988tl,Kuznetsov2009en} while the only solution for MHD shock waves  
was obtained for the case of parallel shock waves propagating along the magnetic field \cite{kvd-osin-pla-2018}. 
Conditions have also been found for the parameters in front of such waves when 
instabilities are generated behind the shock leading to plasma turbulence \cite{kvd-osin-pla-2020}.

The evolutionarity of shock waves in collisionless plasma was studied within the framework of the Chew-Goldberger-Low (CGL) model of anisotropic magnetohydrodynamics in  \cite{Lynn1967,  Morioka1968, Neubauer1970, Zakharov1989}.
Below we will consider the evolutionarity of MHD shock waves in collisionless plasma with heat fluxes 
(8-moment approximation) which  has not been previously studied and present corresponding evolutionarity diagrams.
In the special case of  shock waves propagating perpendicular to the direction of magnetic field the solution coincides with  
results obtained for the CGL equations \cite{Lynn1967,Morioka1968}.
For parallel shock waves the regions of their evolutionarity are determined and expressed in the form of conditions for the upstream shock wave parameters.

\section{Basic equations}

As the basic system of equations for an anisotropic plasma with heat fluxes we will use equations of the 
8-moment approximation for collisionless plasma in a strong magnetic field \cite{Namikawa1981,kvd-osin-pla-2018}. 
These equations, with the exception of two equations for heat fluxes, can be written in divergent form. 
In particular, the equations of energy and magnetic moment conservation have the following form

\begin{align*}
\frac{\partial}{\partial t}\left(\frac{p_\parallel}{2} +p_\perp +\frac{\rho v^2}{2}+\frac{B^2}{8\pi} \right)  + 
\nabla\cdot\left[ 
\left(\frac{p_\parallel}{2} + 2p_\perp + \frac{\rho v^2}{2}+\frac{B^2}{4\pi}\right) \right.  {\bf v}  \;+  \\
\left.\left(p_\parallel-p_\perp-\frac{B^2}{4\pi}\right){\bf v}_\parallel + \left( q^\parallel + q^\perp \right) \frac{\bf B}{B}
  \right] = 0 .
\end{align*}

\begin{align*}
\frac{\partial}{\partial t}\left(\frac{p_\perp}{B}\right) + \nabla\cdot\left( \frac{p_\perp}{B}{\bf v}+
\frac{q^\perp}{B^2} {\bf B} \right) = 0 .
\end{align*}
where $\rho,v,p_\parallel,p_\perp$ are standard notations for plasma density, velocity, parallel and perpendicular 
pressure with respect to direction of magnetic field ${\bf B}$, ${\bf v_\parallel} = ({\bf v}\cdot {\bf B}){\bf B}/B^2$  -- 
longitudinal plasma velocity   and  $q^\parallel, q^\perp$ - fluxes of parallel and perpendicular 
thermal energy along the magnetic field {\bf B} 
\begin{equation*}
q^\parallel =  \int  \frac{m\nu_\parallel^2}{2} \nu_\parallel f(\bm{\nu}) \;d^3\nu \quad , 
\quad q^\perp = \int \frac{m\nu_\perp^2}{2} \nu_\parallel f(\bm{\nu}) \;d^3\nu 
\end{equation*}
where $\bm{\nu}$ is chaotic part of the plasma particle velocity, $f(\bm{\nu})$ is the distribution function with 
even part in $\nu_\parallel$ being bi-Maxwellian. 

The Rankine–Hugoniot (RH) boundary conditions at the shock wave front in an anisotropic plasma with 
heat fluxes follow from the integral equations of  the mass, momentum, energy and magnetic moment conservation,
supplemented with relations following from the Maxwell's equations - continuity of the normal component of the 
magnetic field $B_n$ and the tangential component of the electric field ${\bf E}_\tau$ 
\begin{equation}\label{rh-m}
m = [\rho v_n] = 0  
\end{equation}
\begin{equation}\label{rh-momeq-1}
\left[ p_\perp +\frac{m^2}{\rho} + \frac{B_\tau^2}{8\pi}  + \frac{B_n^2}{B^2} (p_\parallel-p_\perp)  \right] = 0 
\end{equation}
\begin{equation}\label{rh-momeq-23}
\left[ m {\bf v}_\tau  + \left( p_\parallel-p_\perp -\frac{B^2}{4\pi} \right) \frac{B_n{\bf B}\tau}{B^2} \right]=0
\end{equation}
\begin{multline}
m \left[ U+ \frac{m^2}{2\rho^2} + \frac{p_\perp}{\rho} + \frac{v_\tau^2}{2}    +\frac{B_n^2}{B^2} \frac{(p_\parallel-p_\perp)}{\rho} +
\frac{B_\tau^2}{4\pi\rho}  \right] +  \\
B_n \left[   \frac{p_\parallel-p_\perp-B^2/(4\pi)}{B^2} ({\bf B}_\tau\cdot {\bf v}_\tau)  + \frac{q^\parallel+q^\perp}{B}    \right] = 0 
\end{multline}
\begin{equation}\label{rh-pperp}
\left[  \frac{p_\perp }{B} v_n +  \frac{q^\perp}{B^2} B_n \right] =  0 
\end{equation}
 \begin{equation}\label{rh-maxwell}
 [B_n] = 0 \quad, \quad B_n[{\bf v}_\tau] = m \left[\frac{{\bf B}_\tau}{\rho}\right] 
\end{equation}
where $B_n$, ${\bf B}_\tau$, $v_n$ и ${\bf v}_\tau$ - correspondingly normal and tangential components of 
magnetic field and velocity, brackets denote a jump in magnitude across the shock: $[x] = x_2-x_1$. 

The system of  RH relations (\ref{rh-m})-(\ref{rh-maxwell}) is not closed since  nine scalar equations are not enough to 
determine eleven unknown quantities behind the shock given the eleven quantities in front of the shock even if  
shock velocity $D_n$ is known. The above system of relations should be supplemented with two more relations 
including heat fluxes $q^\parallel$ and $q^\perp $. The corresponding conservation laws at the discontinuity are generally unknown but to study the evolutionarity of shock waves, following \cite{Baranov1972en}, it is sufficient to assume that they have 
the form of a general functional relationship between the parameters of plasma before and after the shock front 
and consider various cases of such dependence which affect evolutionarity. These additional relations $F_1$ and $F_2$   
can be written as

\begin{equation}\label{rh-f1}
F_1(\rho_1,p_{\parallel 1},p_{\perp 1},q^\parallel_1,q^\perp_1,{\bf v_1},{\bf B_1},
\rho_2,p_{\parallel 2},p_{\perp 2},q^\parallel_2,q^\perp_2,{\bf v_2},{\bf B_2}, D_n) = 0 
\end{equation}
\begin{equation}\label{rh-f2}
F_2(\rho_1,p_{\parallel 1},p_{\perp 1},q^\parallel_1,q^\perp_1,{\bf v_1},{\bf B_1},
\rho_2,p_{\parallel 2},p_{\perp 2},q^\parallel_2,q^\perp_2,{\bf v_2},{\bf B_2}, D_n) = 0 
\end{equation}
where $D_n$ is the velocity of the shock wave in the laboratory frame of reference, the indices $1,2$ correspond to 
the values of the quantities ahead and behind the shock front. 

In the case of a parallel shock wave, when the magnetic field drops out of boundary conditions, 
the energies associated with longitudinal and transverse degrees of freedom are separately conserved,  
equations for the fluxes of longitudinal  ($q^\parallel$) and transverse ($q^\perp$)  thermal energies 
along the magnetic field take conservative form  and can be used instead of (\ref{rh-f1})-(\ref{rh-f2}) to obtain 
the solution of RH relations \cite{kvd-osin-pla-2018}.

\section{Counting the outgoing waves}

It is well known that in MHD requirement of entropy increase across the shock front is not enough to
select stable, physically realizable (evolutionary) solutions.
Evolutionarity means that the problem of small perturbation of the flow on both sides of the shock and 
perturbation of the shock front itself can be uniquely resolved.

To carry out the evolutionarity analysis, we use the method of counting the number of disturbances (small amplitude waves) 
escaping from the front of discontinuity \cite{Akhiezer1959, Kontorovich1959, Syrovatskii1959, Akhiezer1975, KulikovskiyMHD1965, PolovinDem1990}.
Linearizing 11 boundary conditions (\ref{rh-m})-(\ref{rh-f2}) with respect to small perturbations 
(including the perturbation of the shock velocity $\delta D_n$), 11 equations are obtained relating the perturbed
MHD quantities ahead of the front 
($\delta\rho_1,\delta p_{\parallel 1},\delta p_{\perp 1},\delta q^\parallel_1,\delta q^\perp_1,\delta{\bf v_1},\delta{\bf B_1}$) with
quantities behind the front  
($\delta\rho_2,\delta p_{\parallel 2},\delta p_{\perp 2},\delta q^\parallel_2,\delta q^\perp_2,\delta{\bf v_2},\delta{\bf B_2}$) --
23 quantities in total, 11 on each side of the shock front plus shock velocity perturbation $\delta D_n$.
Excluding the shock velocity perturbation $\delta D_n$ using (\ref{rh-m}) from the linearized equations, 
as well as $\delta B_x$ using the equation for the normal component of the magnetic field (\ref{rh-maxwell}), 
a system of nine linear equations is obtained relating twenty quantities,  the amplitudes of small perturbations
\begin{align*}
\delta\rho_1,\delta p_{\parallel 1},\delta p_{\perp 1},\delta q^\parallel_1,\delta q^\perp_1,\delta v_{x1},\delta v_{y1},\delta v_{z1},\delta B_{y1},\delta B_{z1}, \\
\delta\rho_2,\delta p_{\parallel 2},\delta p_{\perp 2},\delta q^\parallel_2,\delta q^\perp_2,\delta v_{x2},\delta v_{y2},\delta v_{z2},\delta B_{y2},\delta B_{z2} 
\end{align*}
Taking into account the coplanarity theorem \cite{Colburn1966,Chao1970,Lepping1971}, which is valid in our 
case due to the fact that equations (\ref{rh-momeq-23}) and (\ref{rh-maxwell})  remain identical to CGL,
the frame of reference  and  the orientation of the $yz$ axes can be chosen so that on both sides of the 
front $D_n = 0, B_z = 0, v_z = 0$ (the magnetic field and plasma velocity both lie in the $xy$ plane), 
the $x$ axis is normal to the shock front.

Small disturbances disrupt the equilibrium state of the shock front resulting in MHD waves diverging in both directions. If the RH boundary conditions at the shock front make it possible to unambiguously determine the amplitudes of outgoing waves, then the discontinuity is evolutionary. From nine equations it is possible to uniquely determine the amplitudes of nine diverging waves (the initial disturbance evolution problem  has a unique solution).
Otherwise, if the number of outgoing waves is greater or less than the number of equations, there are no solutions or there are infinitely many of them, the discontinuity is non-evolutionary and splits into disturbances of finite amplitude (the initial assumption of the smallness of disturbances is invalid). Thus, the problem of evolutionarity comes down to counting the number of linear waves leaving the front of the shock wave and comparing it with the number of equations (linearized RH relations).

\section{Linear waves} 

The study of  small amplitude (linear) waves in a collisionless plasma with heat fluxes using system of 
equations  of the 8-moment (zero Larmor radius) approximation gives five types of linear waves ($a_i^\pm$) 
propagating in both directions \cite{Namikawa1981,Barkhudarov1987tl,Zakharov1988tl}. The full tenth-order 
dispersion relation in this approximation can be factorized to give the Alfven (transverse) mode which does not 
depend on heat fluxes ($a_A^+ = a_A^{-}$)
\begin{equation}
a_A^2 = V_A^2(1-(\beta_\parallel-\beta_\perp)/2)\cos^2\theta.
\end{equation}
where $V_A^2 = B^2/4\pi\rho$, $\beta_{\parallel,\perp} = 8\pi p_{\parallel,\perp}/B^2$, $\theta$ - the angle between 
vectors ${\bf k}$ and ${\bf B}$,  and four asymmetric  (with respect to the direction of heat fluxes, $a_i^+ \ne a_i^{-}$) 
magnetoacoustic waves related by the general dispersion equation of the eighth-order – fast magnetoacoustic wave 
$F(a_f^\pm)$, slow magnetoacoustic wave $S(a_s^\pm)$, and two intermediate magnetoacoustic waves 
$I_a(a_{Ia}^\pm)$ and $I_b(a_{Ib}^\pm)$, a total of 10 waves in both directions on each side of the shock front, 20 waves 
on both sides of the front in both directions. Phase polars and relative magnitudes of phase velocities for magnetoacoustic 
waves and for the Alfvén wave for specific values of plasma parameters are given in \cite{Namikawa1981}.

\begin{figure}[h]
\includegraphics[width=14cm]{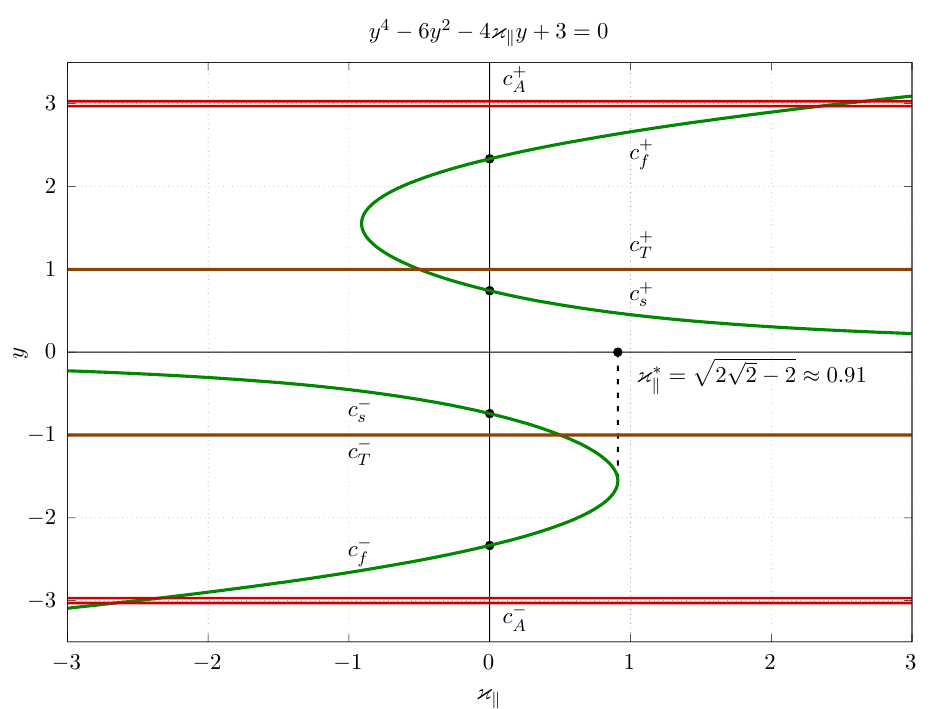}
\caption{Dimensionless  phase velocities $c_i^\pm = a_i^\pm/a_\parallel$ versus dimensionless heat flux  $\varkappa_\parallel$ 
for $c_A^\pm = \pm 3$.} 
\label{fig:ky} 
\end{figure}

For parallel propagation ($\theta = 0$), the dispersion equation for magnetoacoustic modes can be factorized, 
giving three different modes - a transverse (symmetric) mode, the phase velocity of which coincides with that of 
the Alfven wave
\begin{equation}
a_A^2 = V_A^2(1-(\beta_\parallel-\beta_\perp)/2).
\end{equation}
two asymmetric (with respect to the direction of heat fluxes) modes, into which CGL acoustic and entropy modes 
\cite{AbrahamS1967prop} 
convert due to the heat flux $q^\parallel$,  given by the dispersion equation 
\cite{Namikawa1981,kvd-osin-pla-2018,kvd-osin-pla-2020} (see Fig.~\ref{fig:ky})
\begin{equation}\label{dispeq-y}
y^4-6y^2-4\varkappa_{\parallel}y+3=0 .
\end{equation}
where $y=(\omega/ka_\parallel)$, $a^2_\parallel=p_\parallel/\rho$, $\varkappa_\parallel = 2q^\parallel/(p_\parallel a_\parallel)$
and the third, incompressible ``thermal'' ($\delta p_\perp,\delta q^\perp$) mode - the CGL entropy wave \cite{AbrahamS1967prop} modified by the heat flux $q^\perp$, with phase velocity \cite{Namikawa1981}
\begin{equation}
a_T^2 = a_\parallel^2 .
\end{equation}
From equation (\ref{dispeq-y}) it follows that the mode antiparallel to heat flux direction is unstable when
$|\varkappa_\parallel| > \varkappa^*_\parallel = \sqrt{2\sqrt{2}-2} \approx 0.91 $
\cite{Namikawa1981,Zakharov1988tl,kvd-osin-pla-2018}.

\section{Evolutionarity. General case}

In the CGL theory, as well as in the usual MHD, there are two 
out of seven linearized RH relations which (in the special frame of reference) contain only  $\delta v_z$ and $\delta B_z$ 
related to Alfvén perturbations and five remaining linear equations containing other quantities but not  
$\delta v_z$ or $\delta B_z$. 
Same property is true for the system under consideration. The linearized equations (\ref{rh-m})-(\ref{rh-maxwell}) 
of the system of RH relations at the discontinuity split into two independent subsystems - equations 
(\ref{rh-momeq-23}) and (\ref{rh-maxwell}) for the amplitudes of disturbances $\delta v_z, \delta B_z$ in Alfven waves 
(they do not include perturbation of the discontinuity velocity $\delta D_n$ and they are independent of other equations) 
and equations for the amplitudes of disturbances of other quantities related to magnetoacoustic waves. 
The evolutionarity of the discontinuity in this case is given by the overlapping evolutionarity conditions with respect to 
magnetoacoustic and Alfvén perturbations  \cite{Syrovatskii1959}.

Since linearized equations (\ref{rh-f1})-(\ref{rh-f2})  obviously contain quantities $\delta q^\parallel, \delta q^\perp$ 
(perturbations of heat fluxes) and therefore cannot contain values $\delta v_z$ and $\delta B_z$ alone (or only one 
of them), related to Alfvén perturbations, the following options for grouping equations  affecting the evolutionarity are 
possible (total number of equations = number of equations for Alfven waves + number of equations for 
magnetoacoustic waves + number of equations for heat fluxes):

1) ($9 = 2 + 5 + 2 = 2 + 7$) Both linearized equations (\ref{rh-f1}) and (\ref{rh-f2}) do not contain quantities 
$\delta v_z$ , $\delta B_z$ related to the Alfvén perturbations. In this case, two equations relate amplitudes of  the 
Alfven waves  while seven more equations - amplitudes of the remaining waves. For the evolutionarity with respect 
to the Alfven waves, two outgoing waves are required. Seven outgoing waves are required for the magnetoacoustic 
waves.

2) ($9 = 2 + 5 + 2 = 9$) Both equations  (\ref{rh-f1}) and (\ref{rh-f2}) or one of them contains quantities 
$\delta v_z$ , $\delta B_z$ or one of them plus other quantities. In this case, nine equations for amplitudes are not 
separated and evolutionarity requires nine waves leaving the front.
A detailed discussion of the above two cases (1,2) for $\theta\ne0$ and corresponding diagrams are given in the Appendix A.

\section{Evolutionarity of parallel shock waves}

For $ \theta = 0$, in the system of nine linear equations ${\bf v }_\tau=0, {\bf B}_\tau=0$, and in addition to 
the two independent equations for (transverse) Alfvén perturbations $\delta v_z$ and $\delta B_z$, there are 
two more independent equations for $\delta v_y $ and $\delta B_y $ corresponding to the degenerate 
transverse magnetoacoustic wave, the phase velocity of which coincides with the Alfvén velocity.
Since the velocities of these waves coincide,  we have the option (9 = 2 + 5 + 2 = 4 + 5) of four
equations for transverse perturbations (for evolutionarity it is necessary to have four outgoing
waves out of eight) and five equations for the remaining quantities (five outgoing waves out of twelve required).

In this particular case the two-parameter solution $Y_\pm(M_1,\varkappa_{\parallel 1})$ of the RH relations was obtained in 
\cite{kvd-osin-pla-2018} 

\begin{equation}\label{Y}
Y_\pm = \frac{1}{2} + \frac{1}{M_1^2} \pm\sqrt{D},\quad D=\frac{M_1^4-8\varkappa_{\parallel 1} M_1+6}{12 M_1^4}
\end{equation}
where $Y=\{u\}=u_2/u_1$ , $M_1=u_1/a_{\parallel 1}$ (``thermal Mach number'') and $\varkappa_{\parallel 1}$ is the 
dimensionless heat flux. 
As has been noted above, the phase velocity of one of the transverse 
magnetoacoustic waves coincides with the Alfven velocity (the double root of the dispersion equation)
and we have four equations for the transverse amplitudes $\delta v_{y,z}$ and $\delta B_{y,z}$ and five 
equations for the remaining disturbances.
The three of  five linear modes, namely fast and slow magnetoacoustic and thermal mode can be identified by 
dimensionless velocities $c_f, c_s, c_T$ ($c_T=\pm1$, see Fig.~\ref{fig:ky}).
Since three waves moving with the stream in front of the shock ($c^+_{f1},c^+_{s1},c^+_{T1}$) are always 
incoming while three similar waves behind the shock front ($c^+_{f2},c^+_{s2}, c^+_{T2}$)  are always outgoing, 
evolutionarity in relation to these waves requires two more (counter-streaming) outgoing waves out of six on both 
sides, three on each side.
For the two outgoing waves we have three different cases - (2/0),(1/1),(0/2) - according to the number of outgoing waves 
in front of/behind the shock. 
As thermal linear wave $c_T$ may be greater ($\varkappa_\parallel<0.5$) or  less  ($\varkappa_\parallel>0.5$) than 
the slow magnetoacoustic wave $c_s$, we  will mark the first option by $t \; (c_T>c_s)$, second by $s\; (c_s>c_T)$.
Thus, there are two options on each side of the shock front, four for each case,  total of twelve different options. 
Finally, as there are two solutions of the RH relations for parallel shocks ($Y_+$ and $Y_-$  shocks), total number 
of different cases is 24 :
\begin{equation}\label{evo}
\begin{aligned}
(tt/2/0)&:  a_{s1}^- < u_1  < a_{T1}^- < a_{f1}^-\;,\;   u_2 <  a_{s2}^- <   a_{T2}^- <  a_{f2}^-  \; [Y_-]  \\
(tt/1/1)&:  a_{s1}^- < a_{T1}^- <  u_1 <  a_{f1}^-  \;,\;  a_{s2}^- <  u_2<  a_{T2}^-  <  a_{f2}^-\\
(tt/0/2)&: a_{s1}^- < a_{T1}^- < a_{f1}^- <  u_1 \;,\;  a_{s2}^- < a_{T2}^- < u_2 < a_{f2}^-  \; [Y_+]  \\
\\
(st/2/0)&:  a_{T1}^-<   u_1 <  a_{s1}^- < a_{f1}^- \;,\;  u_2 < a_{s2}^- <  a_{T2}^-  < a_{f2}^-  \\
(st/1/1)&:  a_{T1}^- <  a_{s1}^- < u_1 < a_{f1}^-  \;,\; a_{s2}^- <  u_2 < a_{T2}^- < a_{f2}^-  \\
(st/0/2)&:  a_{T1}^-< a_{s1}^- < a_{f1}^- <  u_1  \;,\; a_{s2}^- < a_{T2}^- < u_2 < a_{f2}^-  \; [Y_+] \\
\\
(ts/2/0)&: a_{s1}^-< u_1 <  a_{T1}^- < a_{f1}^-  \;,\; u_2 < a_{T2}^-< a_{s2}^- < a_{f2}^- \;[Y_-]\\
(ts/1/1)&:  a_{s1}^-< a_{T1}^-  <  u_1 < a_{f1}^- \;,\; a_{T2}^-  < u_2 < a_{s2}^- < a_{f2}^- \\
(ts/0/2)&: a_{s1}^- < a_{T1}^- < a_{f1}^- < u_1 \;,\; a_{T2}^-< a_{s2}^- < u_2< a_{f2}^- \\
\\
(ss/2/0)&:   a_{T1}^-< u_1< a_{s1}^- < a_{f1}^- \;,\;   u_2 < a_{T2}^-<  a_{s2}^- < a_{f2}^- \\
(ss/1/1)&:   a_{T1}^-< a_{s1}^- < u_1 < a_{f1}^- \;,\;  a_{T2}^-< u_2<a_{s2}^-  <  a_{f2}^-  \;[Y_-] [Y_+]\\
(ss/0/2)&:   a_{T1}^-<a_{s1}^- <a_{f1}^-< u_1 \;,\; a_{T2}^- < a_{s2}^- < u_2 < a_{f2}^- \;[Y_+]
\end{aligned}
\end{equation}

\begin{figure}[!htb]
\centering
\includegraphics[width=12cm]{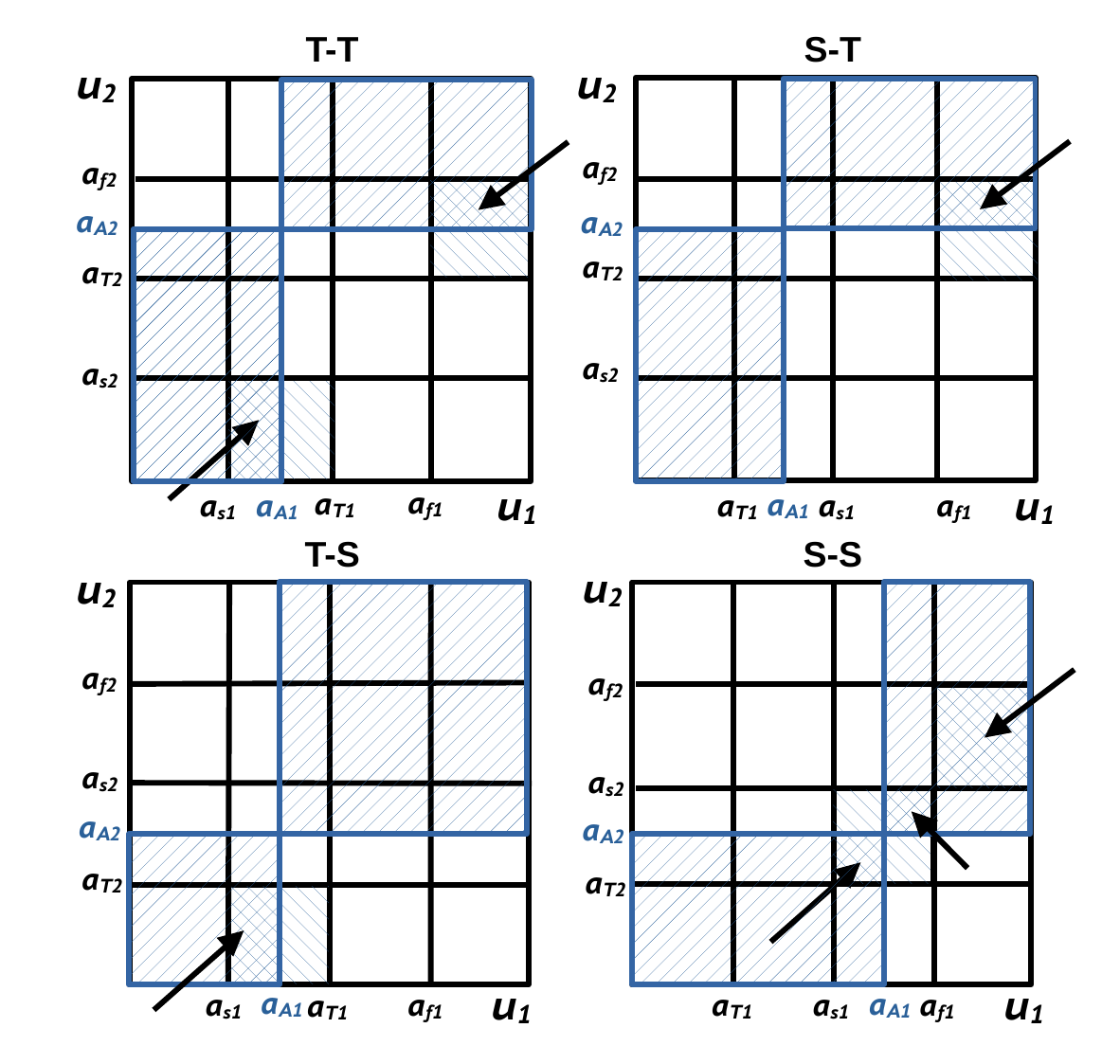}
\caption{Regions of evolutionarity (double hatching) for parallel shock waves on $(u_1,u_2) $ plane }
\label{fig:alf} 
\end{figure}

\begin{figure}[!htb]
\centering
\includegraphics[width=7cm]{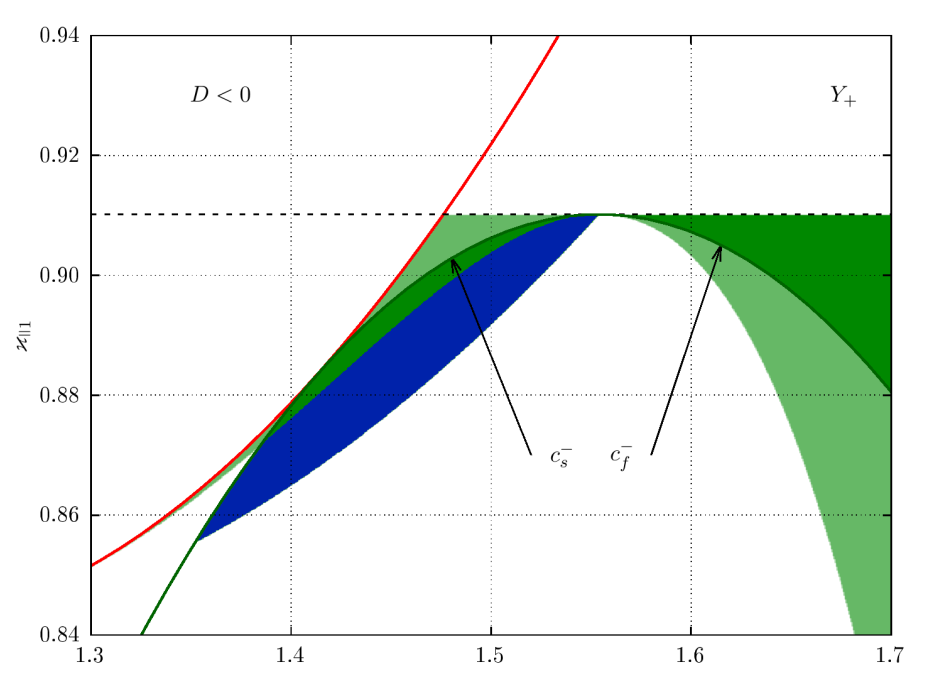}
\caption{Slow  rarefaction evolutionary stable shock waves of $Y_+$ type (thin dark green region near $c_s^-$). 
Dark blue region corresponds to evolutionary rarefaction shock waves with ion-acoustic instability behind the front.}
\label{fig:rshock} 
\end{figure}

The above conditions can also be displayed in the form of diagrams on $(u_1,u_2)$ plane (Fig.~\ref{fig:alf}).
The first condition in each of the above cases  also defines a region on the plane of upstream 
parameters $(M_1,\varkappa_1)$  while  second condition defines a region on the plane of downstream parameters  
$(M_2,\varkappa_2)$. Using the solution (\ref{Y})  of RH relations the reverse 
functions can be used to map the downstream state (and thus the second condition) into the upstream state:  
$(M_2,\varkappa_2) \to (M_1,\varkappa_1)$  \cite{kvd-osin-gna-2020}.
The intersection of these two regions defines a region of evolutionarity although in some cases the intersection may be empty. 
The obtained evolutionarity regions are necessary but not sufficient for an evolutionary shock solution to exist.
Any evolutionary solution should satisfy one of these conditions,  the upstream parameters $(M_1,\varkappa_1)$  
should belong to the intersection of these regions. 
The meaningful solution $Y_\pm(M_1,\varkappa_{\parallel 1})$ also only exists if $D>0, Y>0$ and parallel pressure behind 
the shock is positive, $P=\{ p_\parallel \} > 0$. The stability condition $|\varkappa_{\parallel 1}|<\varkappa_{\parallel}^*$ 
(in front of the shock) must be met so that the surrounding plasma was stable while condition 
$|\varkappa_{\parallel 2}|<\varkappa_{\parallel}^*$ (behind the shock front) is required for the stationary solutions to exist.
Cases with nonempty intersection (evolutionary) are marked in (\ref{evo}) by the $Y_\pm$, corresponding to the type 
of the shock solution and are also marked by the left-slanted hatching on Fig.~\ref{fig:alf}. 
The case $(ss/1/1)$ demonstrates the existence of slow ($M_1\approx c_{s1}^-$) evolutionary stable rarefaction shocks for 
$Y_+$. Slow evolutionary and stable compression $Y_-$ solutions also exist in this region (Fig.~\ref{fig:rshock}).


Since one double (two amplitudes) Alfvén wave behind the shock front is always outgoing, another double outgoing wave is necessary for the evolutionarity with respect to Alfvén waves, either in front or behind the shock front,  
which is determined by the following conditions
\begin{equation}\label{evopar}
\begin{aligned}
&u_1<a_{A1}, u_2< a_{A2} \\
&u_1>a_{A1}, u_2> a_{A2}
\end{aligned}
\end{equation}
The addition of these regions to ($u_1,u_2$) diagrams (Fig,~\ref{fig:alf}, right-slanted hatching) defines combined 
regions of evolutionarity (double-hatching regions).   

As has already been noted, using the solution obtained in \cite{kvd-osin-pla-2018}, the parameters of the plasma 
behind the shock can be expressed in terms of the parameters in front of the shock. In this case, the evolutionarity 
conditions with respect to Alfvén waves (\ref{evopar}) can be explicitly written in the form of 
conditions on the parameters ahead of the front. 
\begin{equation}
M_1^2 \lessgtr A_1^2  \Leftrightarrow 
M_1^2 \lessgtr  \frac{2}{\beta_{\parallel 1}} + \frac{\beta_{\perp 1}}{\beta_{\parallel_1}} -1
\end{equation}
\begin{equation}
M_2^2 \lessgtr A_2^2  \Leftrightarrow 
M_1^2 \lessgtr  \frac{2}{\beta_{\parallel 1} } + \frac{\beta_{\perp 1}}{\beta_{\parallel_1}}  \{p_\perp\}  -1       
\end{equation}
where $M_{1,2} = u_{1,2}/a_{\parallel 1,2}$ ,   $A_{1,2} = a_{A1,2}/a_{\parallel 1,2}$ and $\{p_\perp\} = p_{\perp 2} / p_{\perp 1} $ is the ratio of perpendicular pressure values at the shock front which is expressed through the parameters ahead of the front  \cite{kvd-osin-pla-2018}. 

\begin{figure}[!htb]
\includegraphics[width=7cm]{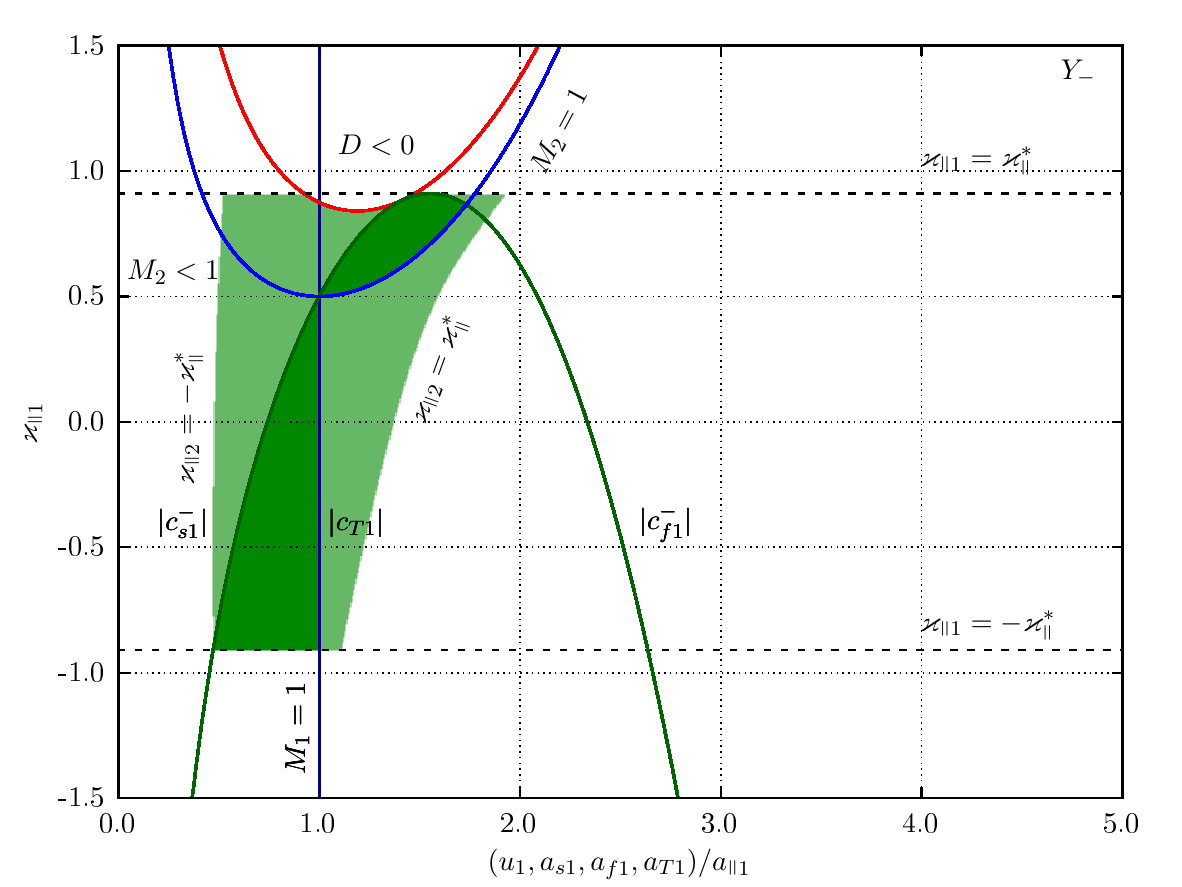}
\includegraphics[width=7cm]{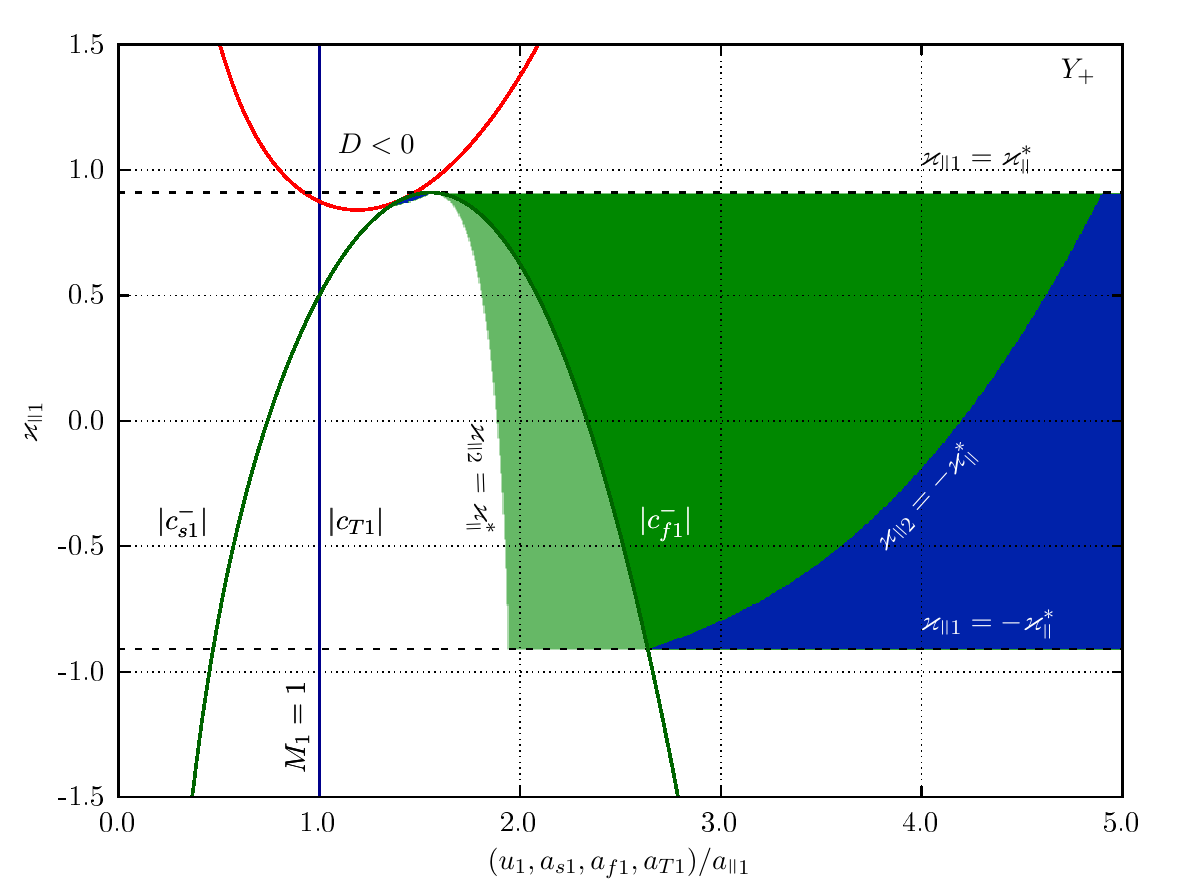}
\caption{Regions of evolutionarity (dark green/blue) in relation to magnetoacoustic perturbations for $Y_-$ and $Y_+$  shock waves. Stable evolutionary shocks are in dark green.}
\label{fig:evol-ms} 
\end{figure}

\begin{figure}[!htb]
\includegraphics[width=7cm]{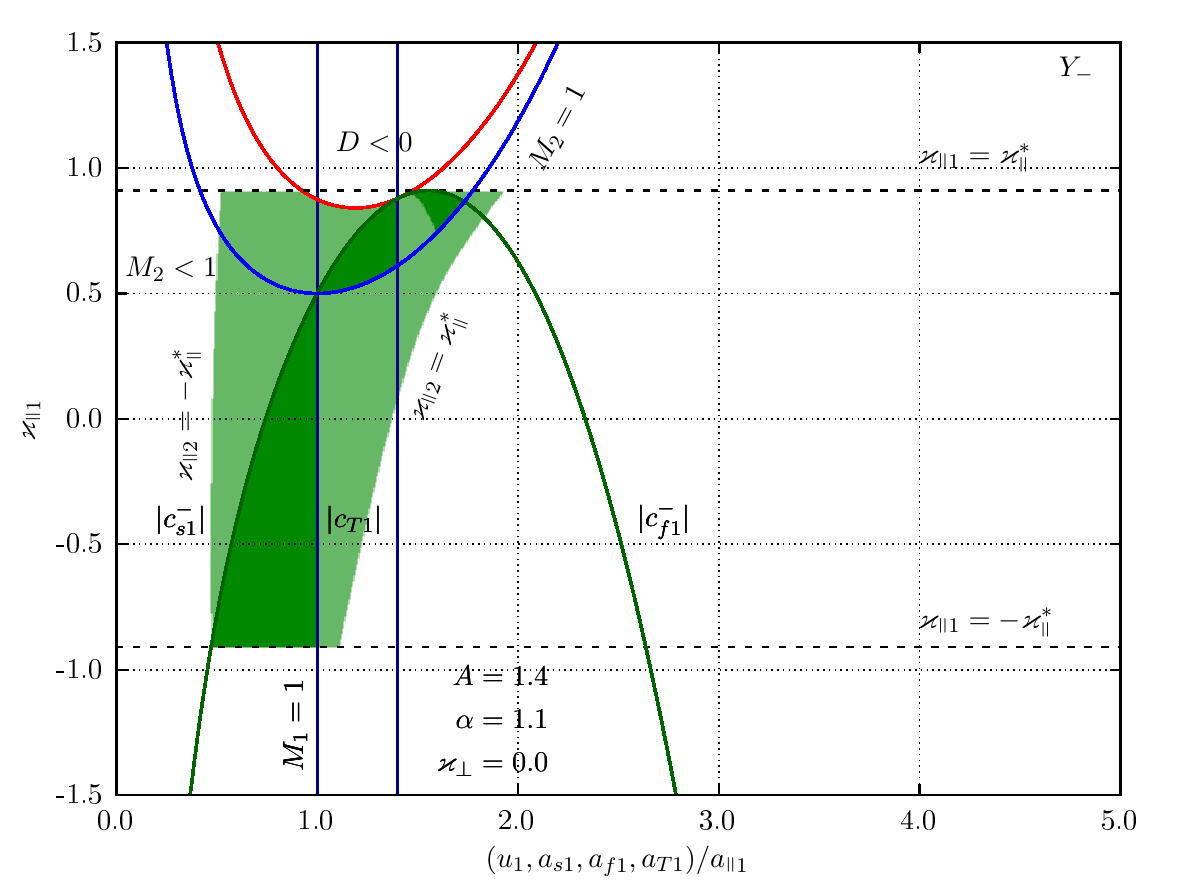}
\includegraphics[width=7cm]{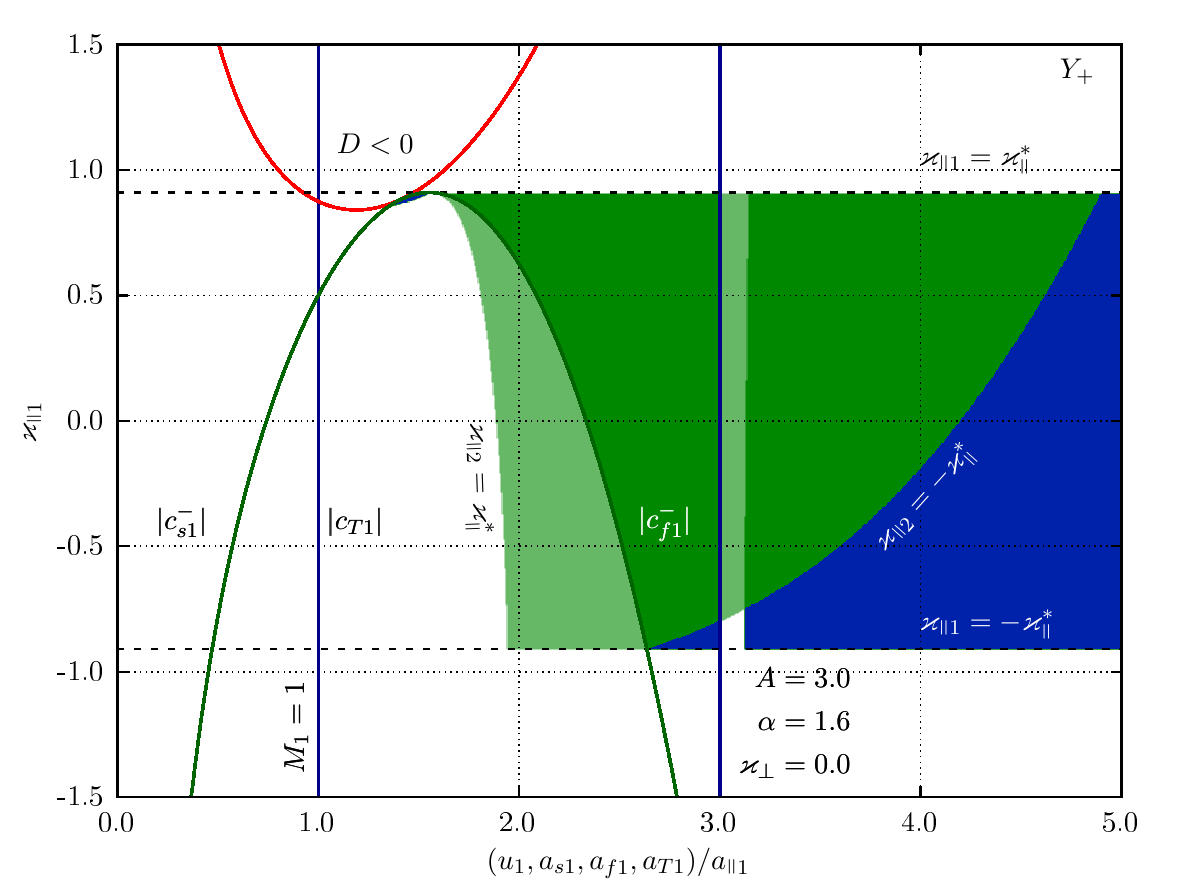}
\caption{Regions of evolutionarity (dark green/blue) in relation to magnetoacoustic and Alfven perturbations for  $Y_-$ and $Y_+$  
shock waves. Stable evolutionary shocks are in dark green.}
\label{fig:evol-a} 
\end{figure}

In Fig.~\ref{fig:evol-ms},\ref{fig:evol-a} the evolutionarity regions for $Y_-$ and $Y_+$ shock waves are shown on the 
parameter plane ($M_1,\varkappa_{\parallel 1}$) separately with respect to magnetoacoustic waves 
(Fig.~\ref{fig:evol-ms}), and with respect to all waves (i.e. including Alfvén modes, Fig.~\ref{fig:evol-a}) when 
$A_1 = 1.4$ for the $Y_-$ shock wave and $A_1=3$ for the $Y_+$ shock wave. 
Areas of evolutionarity are shown in dark green/blue colors.  
For $Y_-$, compression waves ($Y_- < 1$) with $M_1>c_s^- $ and $M_1<1, M_2<1$ or $M_1>1, M_2>1$   
(two areas in dark green color on Fig.~\ref{fig:evol-ms})  are evolutionary and stable, they lie entirely within stability region 
$|\varkappa_{\parallel 1,2}|<\varkappa_{\parallel}^*$ (green). Rarefaction $Y_-$ solutions ($Y_->1$) are not evolutionary.
For $Y_+$, the evolutionarity region for compression waves ($Y_+<1$) is the dark green and dark blue region but 
the latter region is unstable behind the front. 


\section{Results}

Boundary conditions at the shock front are considered for the system of MHD equations describing collisionless 
anisotropic plasma with heat fluxes (8-moment approximation). 
For the boundary conditions related to the heat fluxes, their general functional dependence on variables was 
used and  possible cases of such dependence affecting evolutionarity were considered. 
For several special cases, conditions for the evolutionarity of shock waves are obtained in the form of relations for the 
plasma flow velocity in front of the shock ($u_1$) and behind it ($u_2$) and corresponding diagrams on the 
plane ($u_1, u_2$) are presented. 
The evolutionarity of super-Alfvenic and sub-Alfvenic shock waves, which can have different velocities with respect 
to the phase velocities of magnetoacoustic waves, is shown. 

For parallel shock waves, when one of the magnetoacoustic waves degenerates into a transverse Alfven wave, 
conditions for their evolutionarity are obtained and, using previously obtained solution for jumps in quantities at 
the front of these shock waves, these conditions are expressed in terms of parameters $(M_1,\varkappa_{\parallel 1})$ 
of the plasma in front of the shock. Evolutionarity conditions are satisfied for the fast compression shock waves 
$M_1>c_f$ and for the slow compression shock waves for which flow velocity is below ($M_1<c_{T1}, M_2<c_{T2}$)  
or above ($M_1>c_{T1}, M_2>c_{T2}$) thermal velocity $c_T$ both in front and behind shock front.  
Slow ($M_1 \approx c_s$) rarefaction shock waves are also found to be evolutionary and stable in a restricted 
region of parameters.

\section{Conclusions}

The regions of evolutionarity of MHD shock waves in collisionless plasma with heat fluxes are determined which allow 
for the existence of super-Alfvénic and sub-Alfvénic shock waves. Five linear waves existing in this MHD model for
different values of the magnetic field allows one to classify shock waves as slow, intermediate and fast. 
For the previously found solution for parallel shock wave an overlap of the regions of stability and 
evolutionarity is determined and shown on the plane of parameters in front of the shock front.

\newpage

\appendix

\section{General case ($\theta\ne 0$)}
\label{a1}
Below, when describing magnetoacoustic waves, we use the notation from \cite{Namikawa1981}.

Case 1. ($9 = 2 + 7$).  The evolutionarity condition requires that there were two outgoing Alfvén waves 
and seven outgoing magnetoacoustic waves. Since behind the shock front one Alfven wave is always outgoing, 
the condition for two outgoing Alfven waves correspond to the following conditions

1.1) $u_1<a_{A1}, u_2< a_{A2}$ - second outgoing Alfvén wave ahead of the front 
 
1.2) $u_1>a_{A1}, u_2>a_{A2}$ - second outgoing Alfvén wave behind the front
In the same way, four magnetoacoustic waves behind the shock front propagating downstream, are outgoing, so the 
following inequalities correspond to the existence of three additional outgoing magnetoacoustic waves

1.3) $a_{s1} < u_1 < a_{Ib 1}, u_2 < a_{s2}$ - three outgoing waves ($I_{a1}, I_{b1}, F_1$) in front of the shock, 
no outgoing waves behind the shock (the variant of  3/0, 3 fastest waves in front of the shock).

1.4) $a_{Ib1} < u_1 <  a_{Ia1}, a_{s2} < u_2 < a_{Ib2}$ - two outgoing waves in front  ($I_{b1}, F_1$), one behind the front  ($S_2$) (the variant of 2/1, 2 fastest waves in front, 1 slowest behind the front).

1.5) $a_{Ia1} < u_1 < a_{f1}, a_{Ib2} < u_2 < a_{Ia2}$ - one outgoing wave in front ($F_1$), two outgoing waves behind the front  ($S_2, I_{b2}$) (the variant of 1/2,  1 fastest wave in front, 2 slowest waves behind the front).

1.6) $a_{f1}  < u_1, a_{Ia2} < u_2 < a_{f2} $, - no outgoing waves in front, three outgoing waves behind the front 
($S_2, I_{b2}, I_{a2}$), (the variant 0/3, 3 slowest waves behind the front).

\begin{figure}[!htb]

\includegraphics[width=6.5cm]{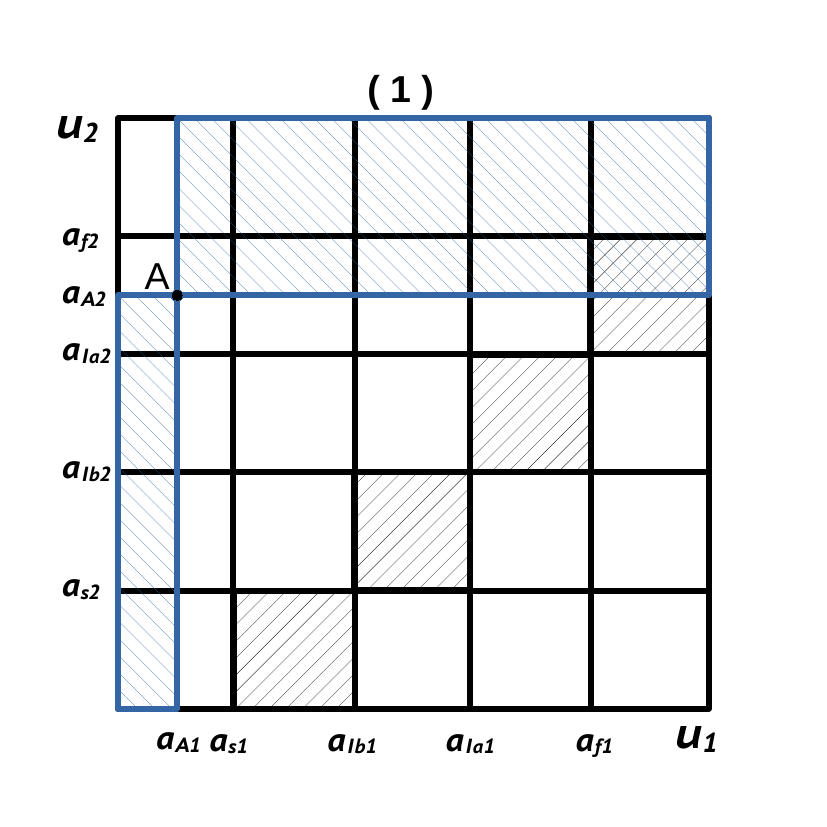}
\includegraphics[width=6.5cm]{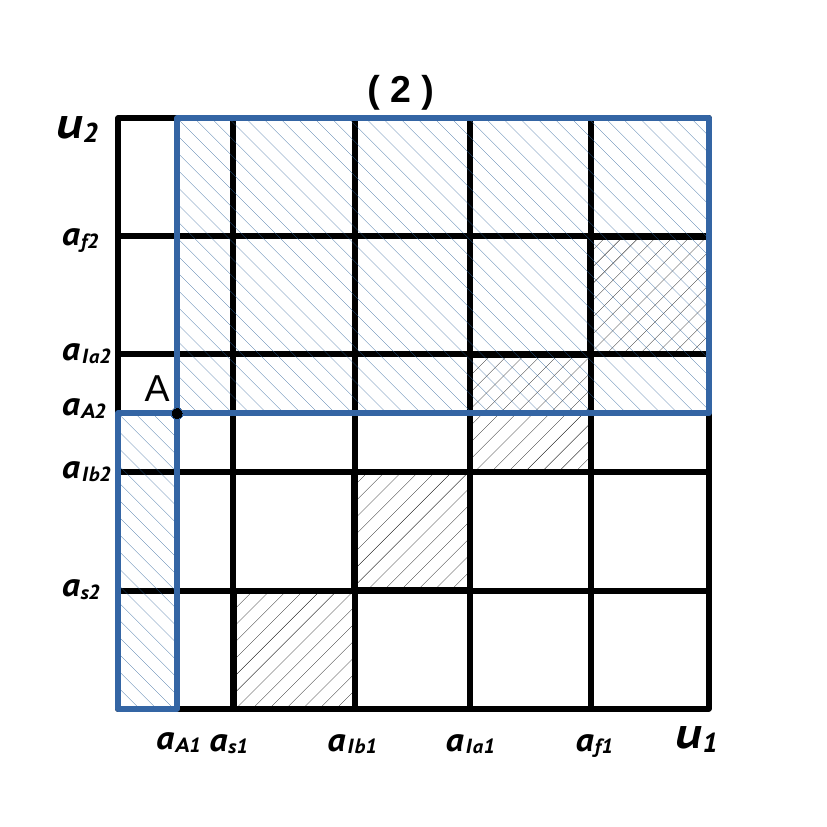}

\includegraphics[width=6.5cm]{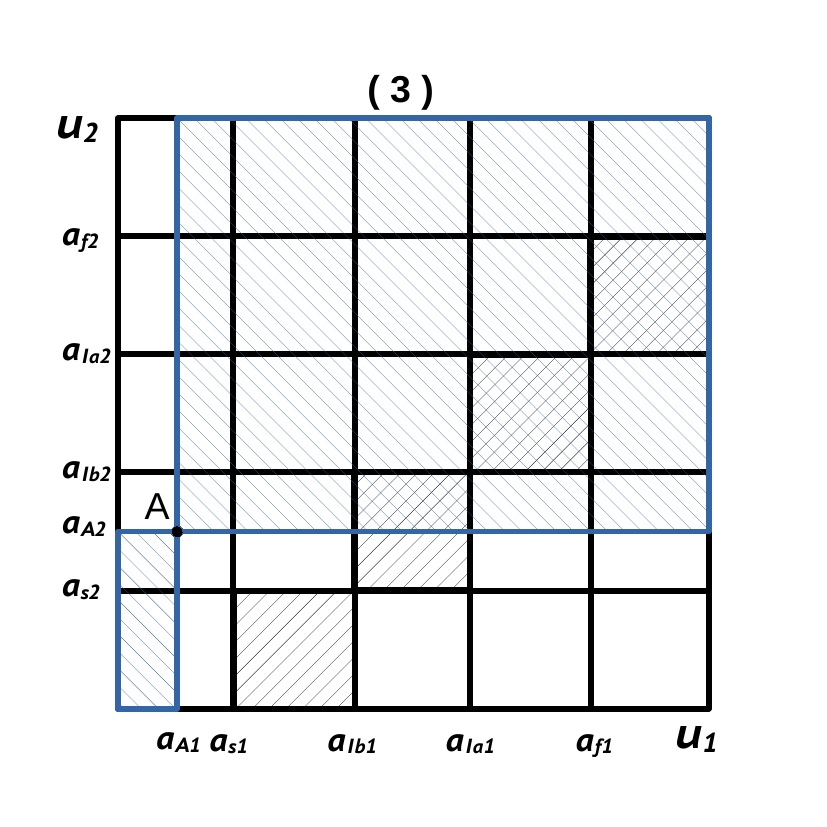}
\includegraphics[width=6.5cm]{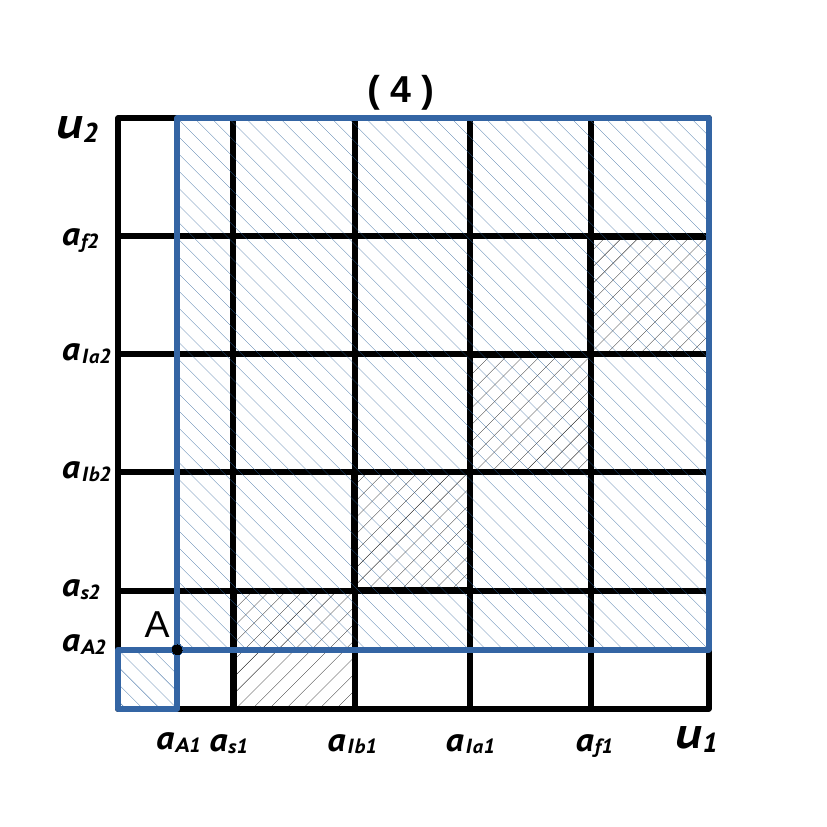}
\caption{1-4}
\label{fig:grid1-4} 
\end{figure}

\begin{figure}[!htb]
\includegraphics[width=6.5cm]{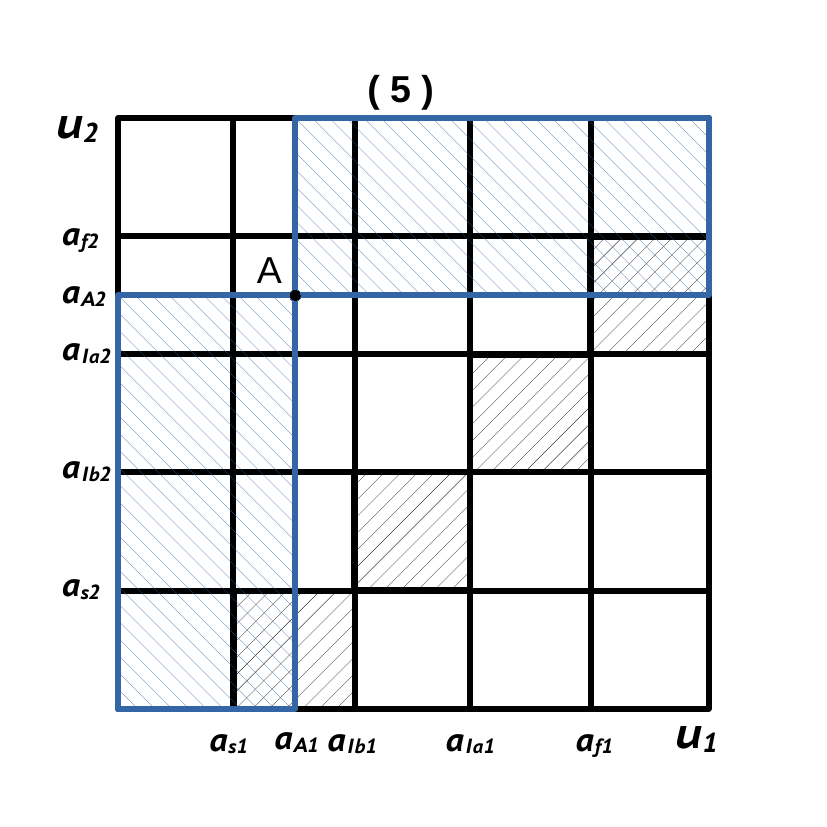}
\includegraphics[width=6.5cm]{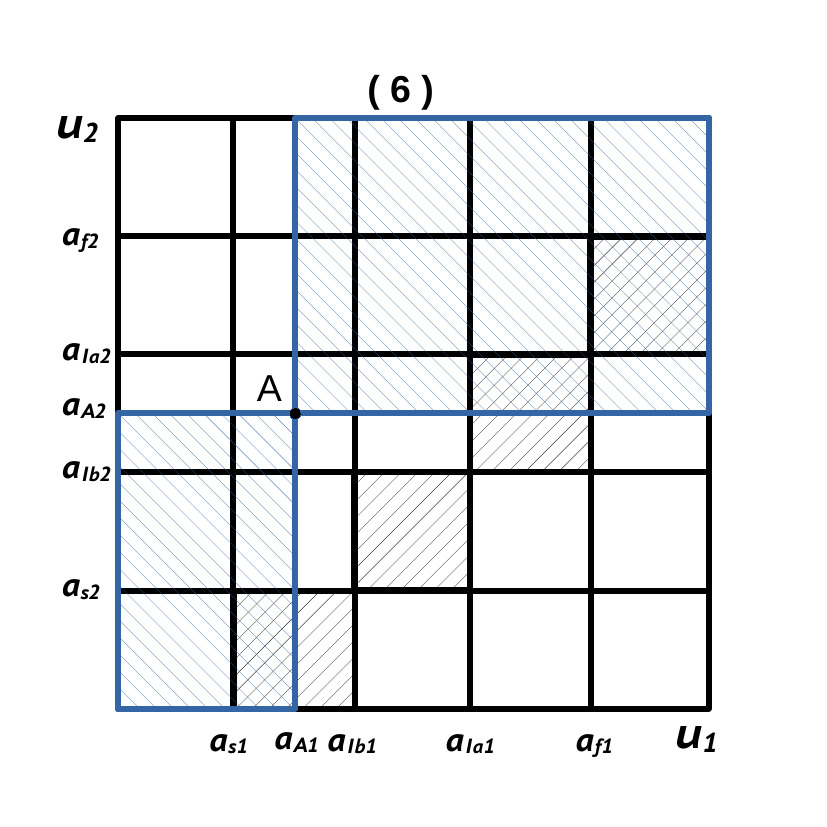}

\includegraphics[width=6.5cm]{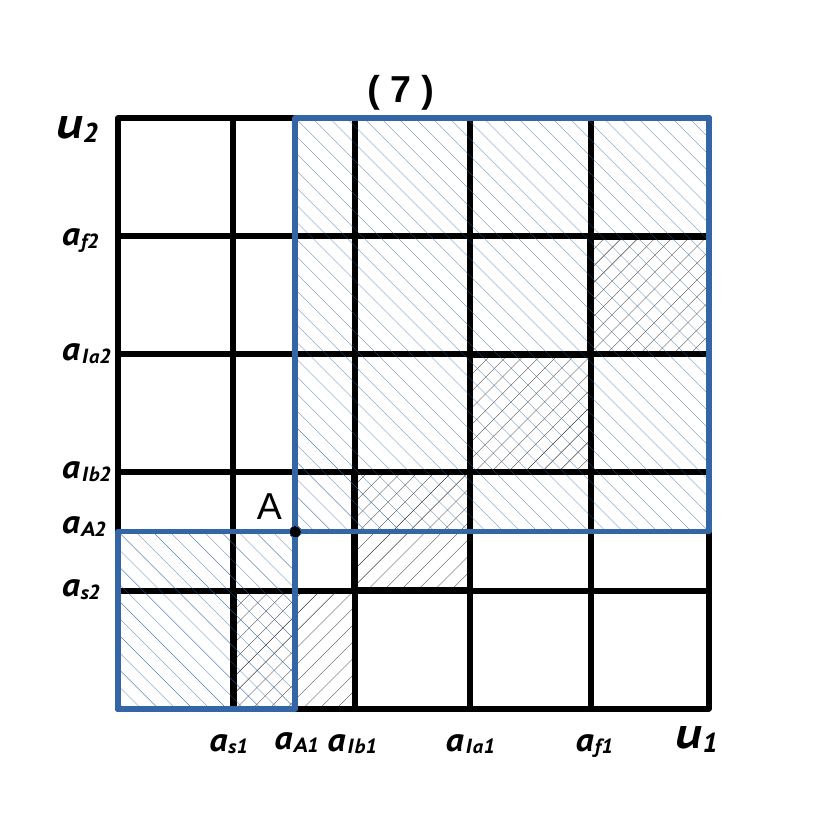}
\includegraphics[width=6.5cm]{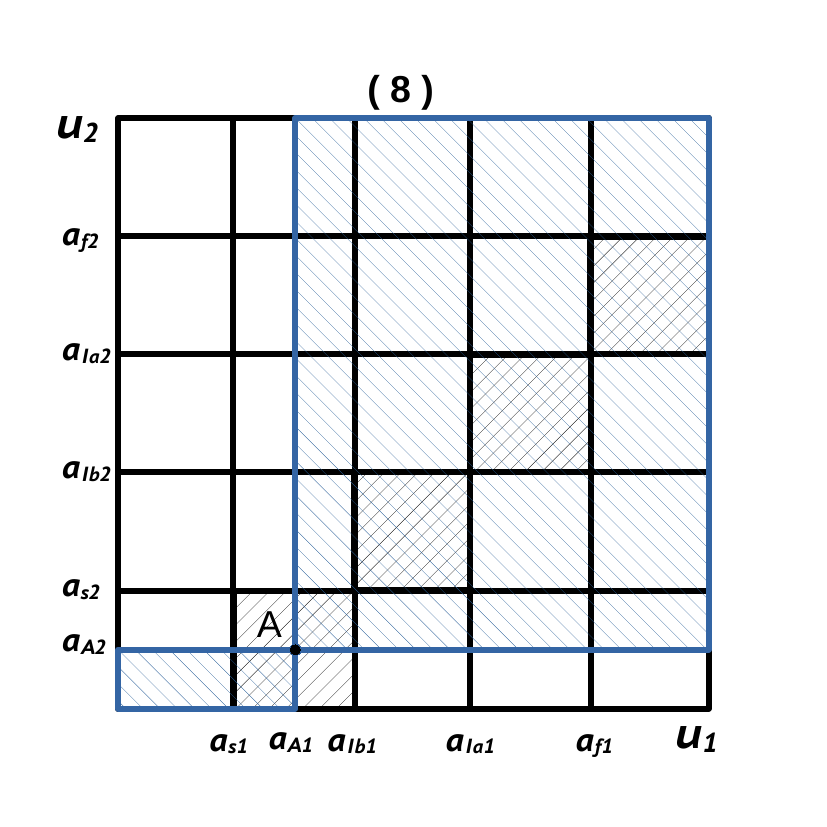}
\caption{5-8}
\label{fig:grid5-8} 
\end{figure}

\begin{figure}[!htb]
\includegraphics[width=6.5cm]{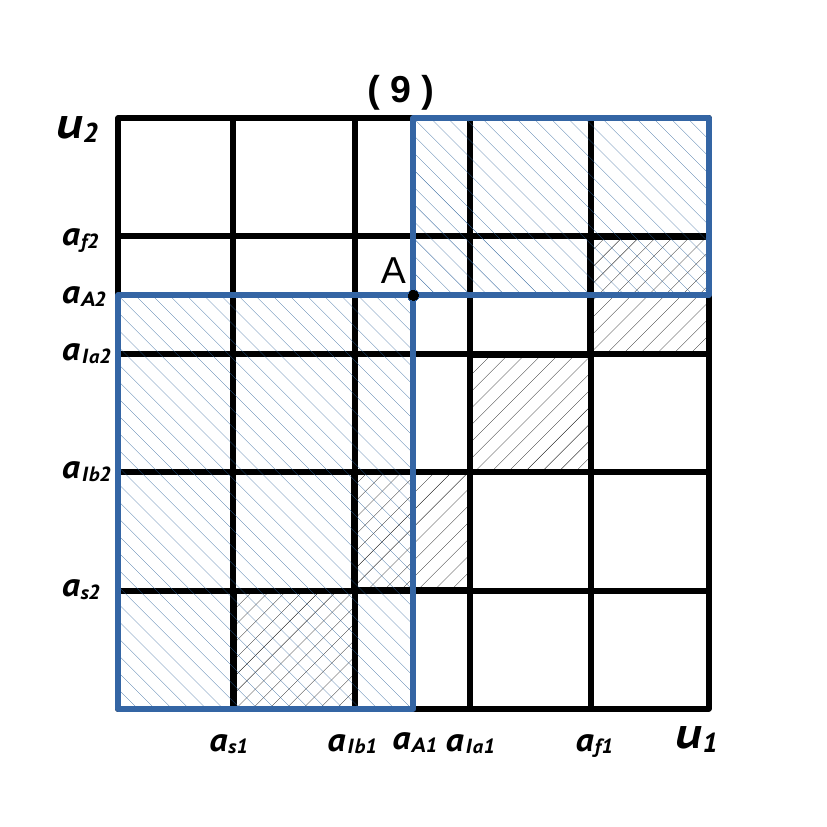}
\includegraphics[width=6.5cm]{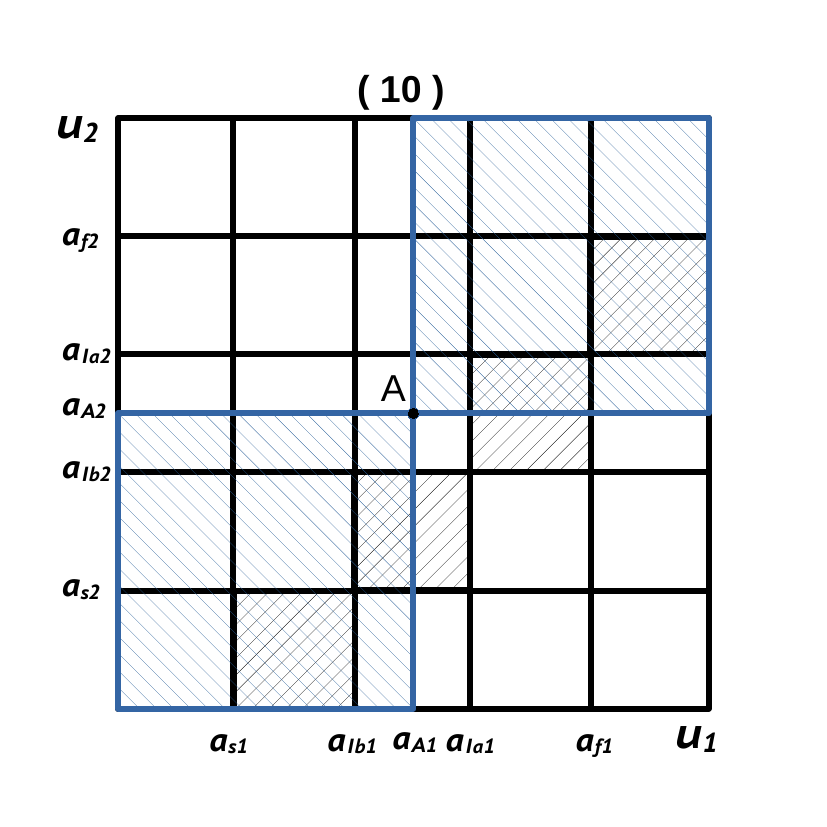}

\includegraphics[width=6.5cm]{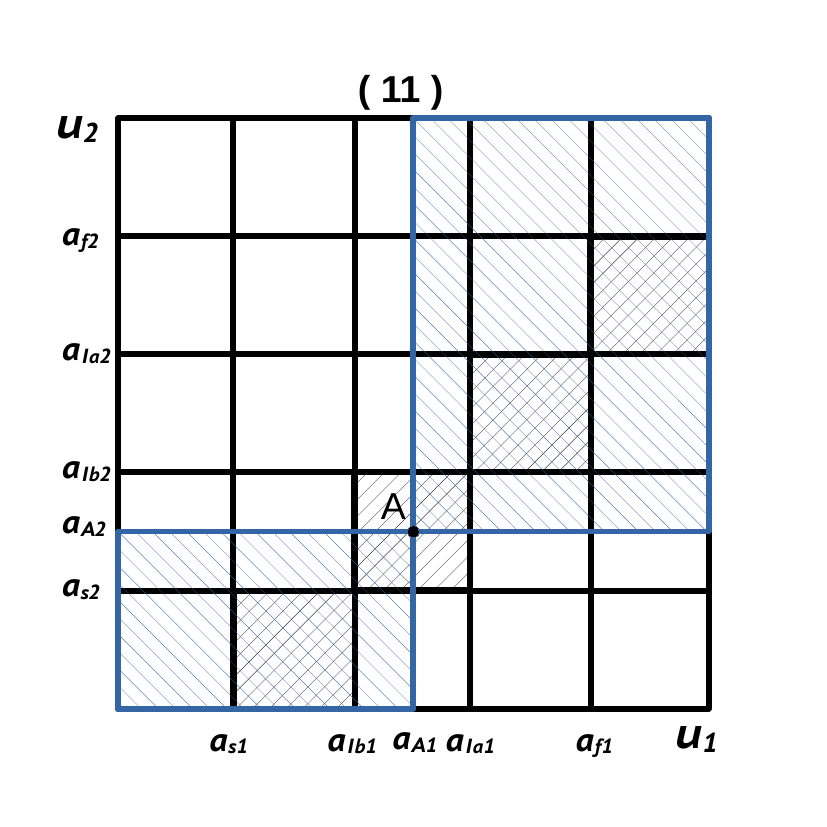}
\includegraphics[width=6.5cm]{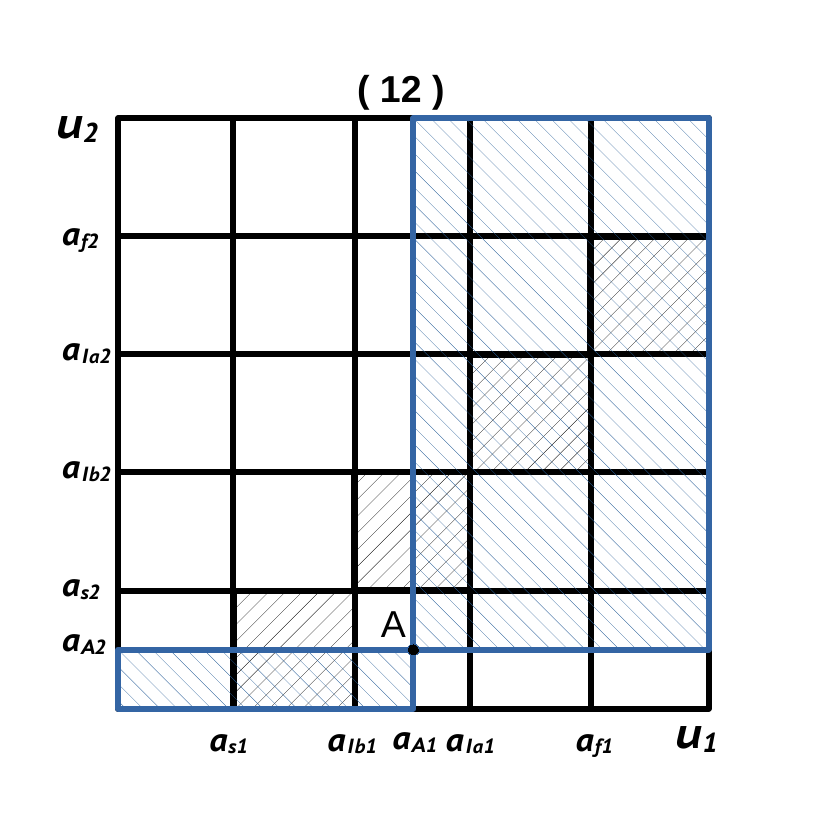}
\caption{ 9-12.}
\label{fig:grid9-12} 
\end{figure}

\begin{figure}[!htb]
\includegraphics[width=6.5cm]{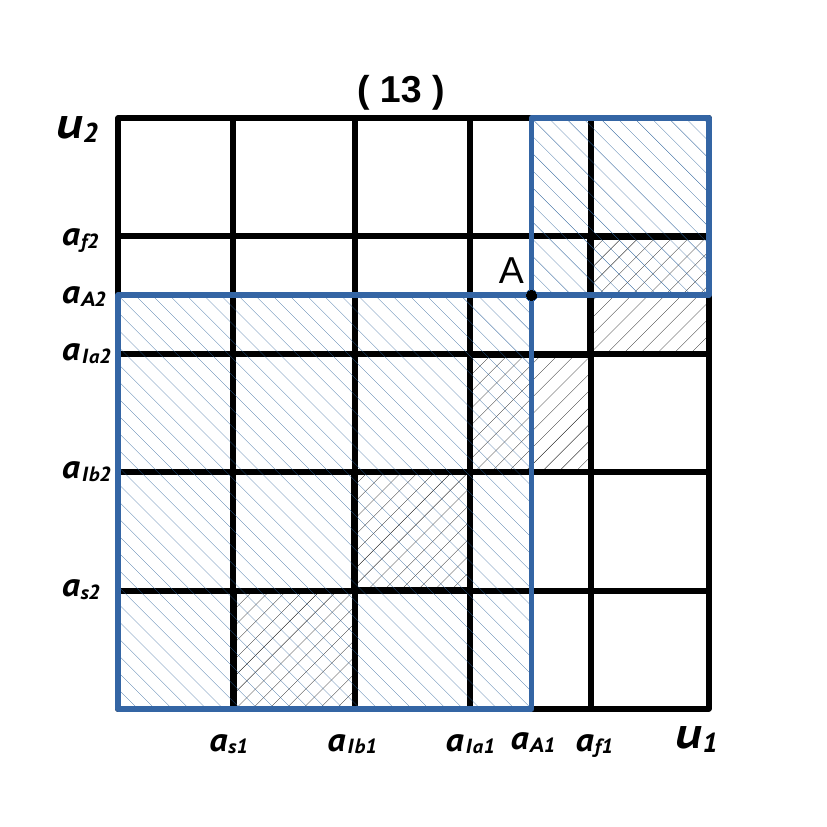}
\includegraphics[width=6.5cm]{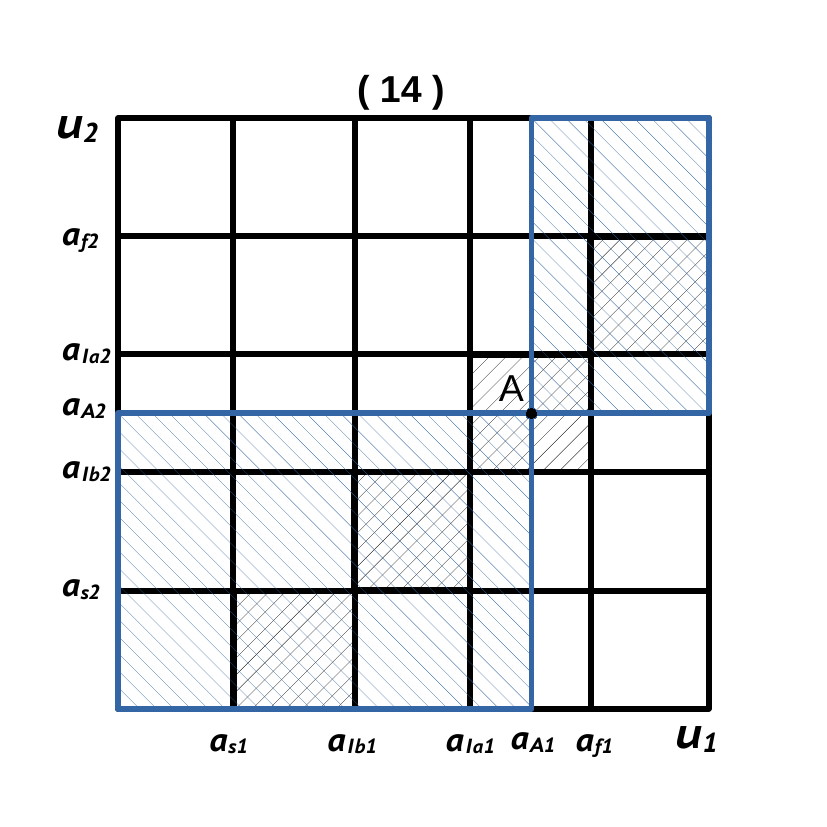}

\includegraphics[width=6.5cm]{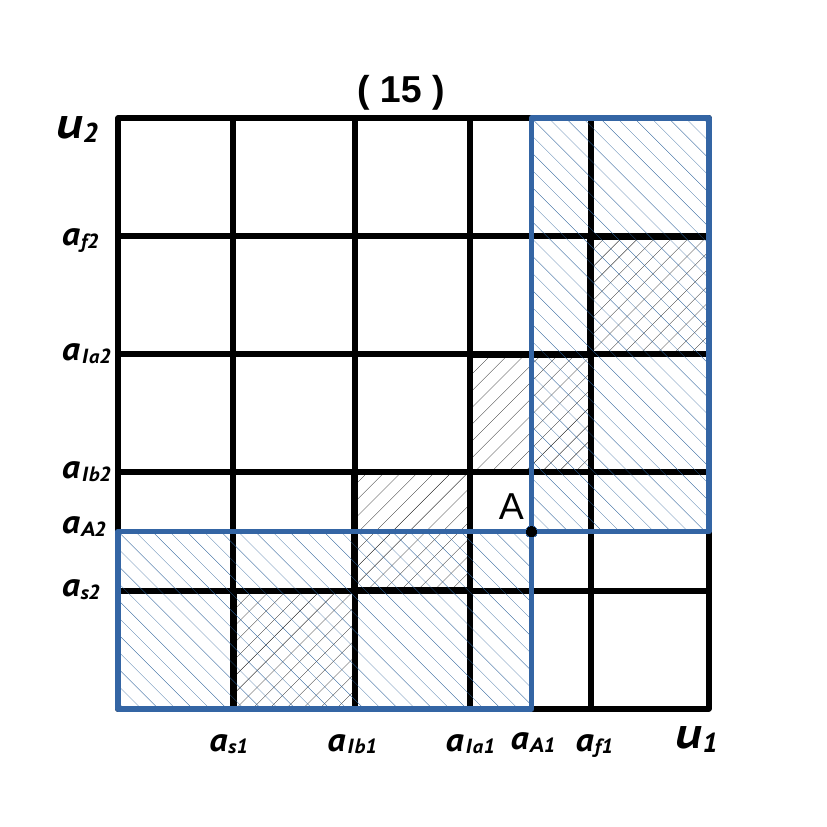}
\includegraphics[width=6.5cm]{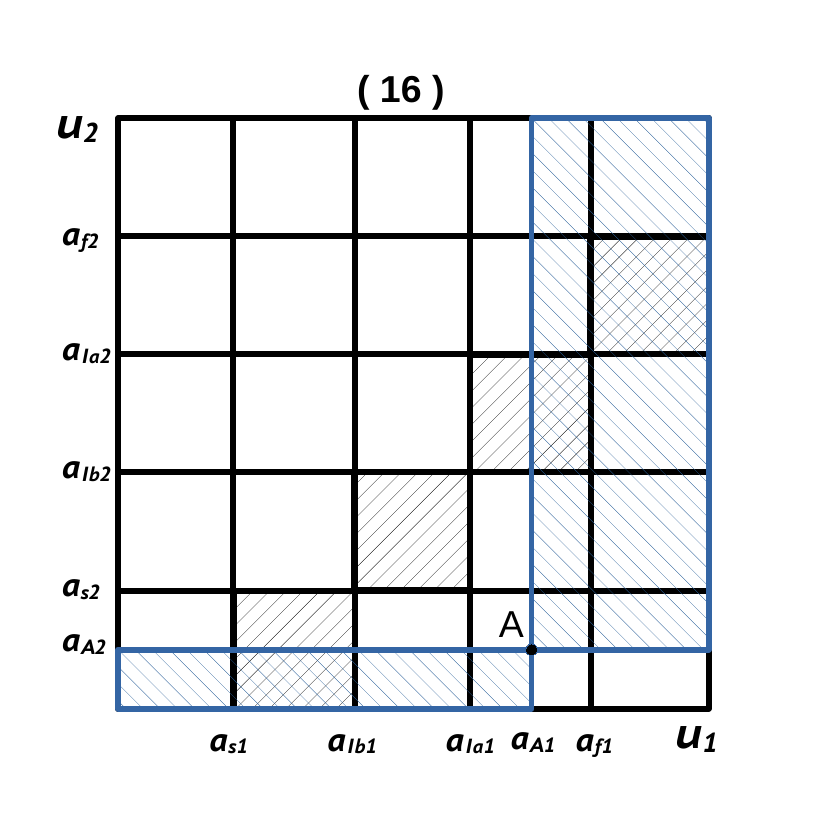}
\caption{ 13-16}
\label{fig:grid13-16} 
\end{figure}

In the diagrams Fig.~\ref{fig:grid1-4},\ref{fig:grid5-8},\ref{fig:grid9-12},\ref{fig:grid13-16} regions of evolutionarity on the plane 
($u_1 ,u_2$) correspond to the regions of intersecting hatchings in the rectangles determined by the above conditions.
Point A defines position of Alfvén velocities in front of the shock wave front and behind it.
In total, taking into account that $a_A \leq a_f$, we have 16 options for the location of point A.
The diagrams give only  general idea of the 56 regions of evolutionarity of the shock wave since wave velocities behind 
the shock front depend on the parameters in front of the shock and in general case such dependence is unknown.

On Fig.~\ref{fig:grid1-4} evolutionarity regions include fast ($u_1 > a_{f1}$) and 
intermediate ($a_{s1} < u_1 < a_{f1} $) superalfvenic shock waves ($u_1 > a_{A1}$), 
On Fig.~\ref{fig:grid5-8},\ref{fig:grid9-12},\ref{fig:grid13-16} – fast and intermediate superalfvenic and intermediate subalfvenic  ($u_1 < a_{A1}$) shock waves.

Case 2. ($9 = 2 + 5+2 = 9$) 
For the evolutionarity, nine outgoing waves are required since  Alfven and magnetoacoustic waves in this case do not separate. 
Five waves behind the front are always outgoing, additional four outgoing waves are needed. Possible options 
are 4/0, 3/1, 2/2, 1/3 and 0/4 (fastest in front of the shock/slowest behind the shock). Since phase 
velocities of magnetoacoustic waves have well-defined relative positions and phase velocity of  the Alfvén wave relative 
to them can take any value provided $a_A \leq a_f$, there are 80 different options that affect the evolutionarity 
(16 options for the position of point A as in Case 1, each having 5 options for outgoing waves - 4/0, 3/1, 2/2, 1/3 and 0/4).
All these cases can be considered by analogy with Case 1 so we will not present all of them here. We shall only present two 
extreme cases - low Alfven velocity in front of the shock and high Alfven velocity behind the front (Case 2.1) and vice versa (Case 2.2).

{\bf Case 2.1.} $a_{A1} < a_{s1}, a_{Ia2} < a_{A2} < a_{F2}$ 

Variant  2.1.1. (4/0) 

$a_{A1} < u_1 < a_{s1}$   four outgoing waves in front  

$u_2 < a_{s2}$         no additional outgoing waves behind the front 

Variant  2.1.2. (3/1) 

$a_{s1} < u_1 < a_{Ib1}$, three outgoing waves in front 

$a_{s2} < u_2 < a_{Ib2}$ one additional outgoing wave behind the front

Variant  2.1.3. (2/2)

$a_{Ib1} < u_1 < a_{Ia1}$  two outgoing waves in front of the shock

$a_{Ib2} < u_2 < a_{Ia2}$  two additional outgoing waves behind the front

Variant   2.1.4. (1/3)

$a_{Ia1} < u_1 < a_{f1}$, one outgoing wave in front

$a_{Ia2} < u_2 < a_{A2}$ three additional outgoing waves behind the front

Variant  2.1.5. (0/4)

$a_{f1} < u_1$   no outgoing waves in front

$a_{A2} < u_2 < a_{f2}$ four additional outgoing waves behind the front

{\bf Case 2.2.} $a_{Ia1} < a_{A1} < a_{f1} , a_{A2} < a_{s2} $

Variant  2.2.1. (4/0)

$a_{s1} < u_1 < a_{Ib1}$, four outgoing waves in front

$u_2 < a_{A2}$     no additional outgoing waves behind the front

Variant  2.2.2. (3/1)

$a_{Ib1} < u_1 < a_{Ia1}$  three outgoing waves in front

$a_{A2} < u_2 < a_{s2}$  one additional wave behind the front 

Variant  2.2.3. (2/2)

$a_{Ia1} < u_1 < a_{A1}$ two outgoing waves in front 

$a_{s2} < u_2 < a_{Ib2}$  two additional outgoing waves behind the front

Variant  2.2.4. (1/3)

$a_{A1} < u_1 < a_{f1}$, one outgoing wave in front 

$a_{Ib2} < u_2 < a_{Ia2}$  three additional outgoing waves behind the front

Variant  2.2.5. (0/4)

$a_{f1} < u_1$     no outgoing waves in front

$a_{Ia2} < u_2 < a_{f2}$ four additional outgoing waves behind the front

\begin{figure}[!htb]
\centering
\includegraphics[width=6.5cm]{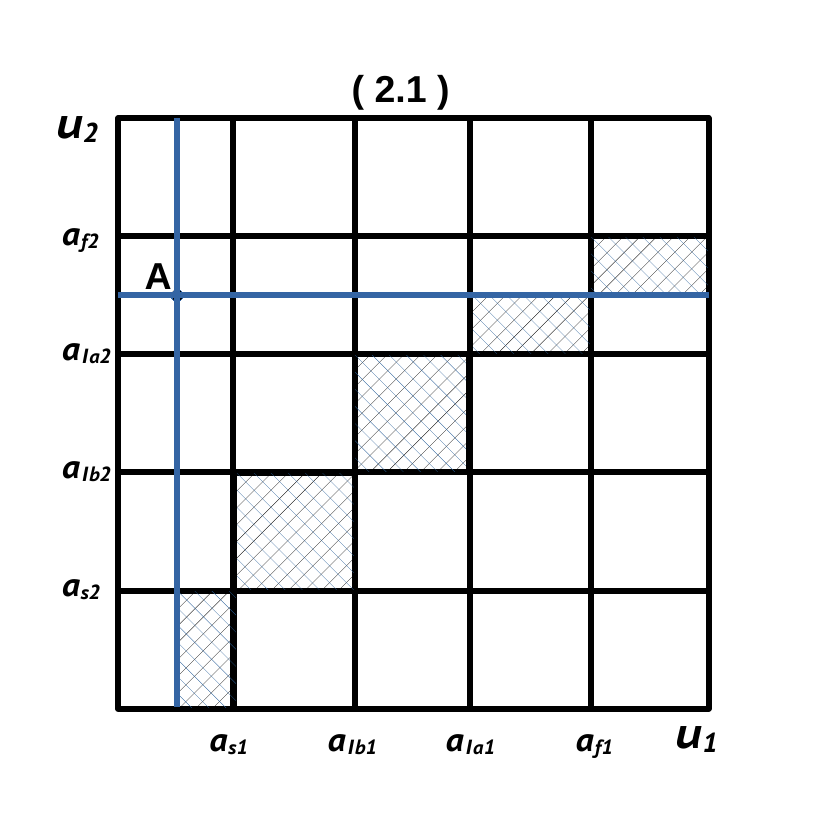}
\includegraphics[width=6.5cm]{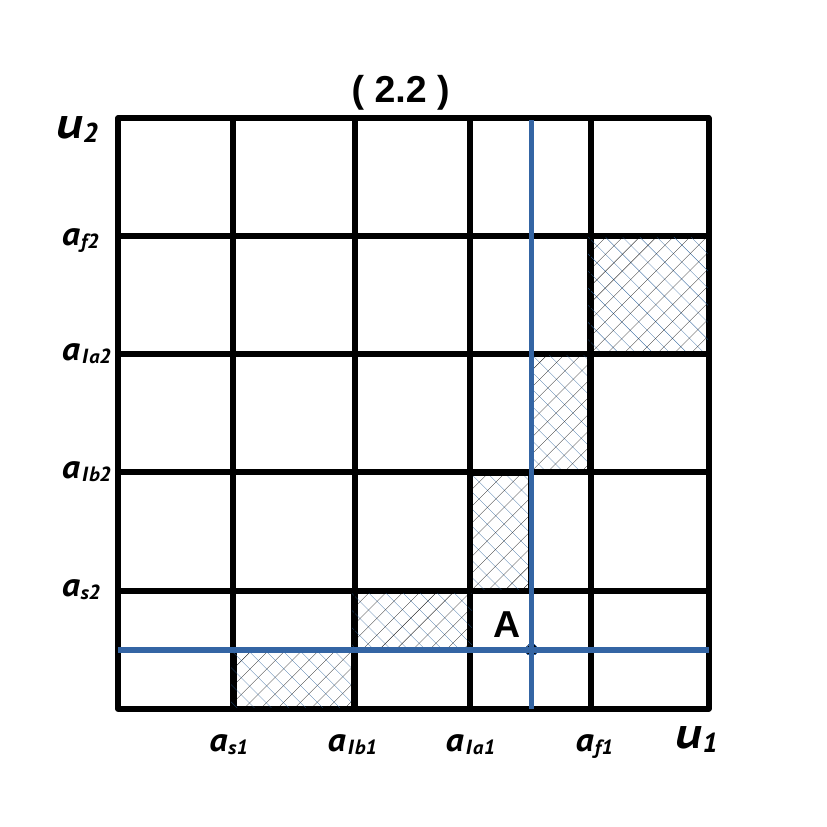}
\caption{Cases 2.1, 2.2}
\label{fig:21} 
\end{figure}

\noindent
On Fig.~\ref{fig:21} double hatching show areas of evolutionarity for the cases 2.1 and 2.2
In the general case, for a shock wave propagating at an arbitrary angle to the background magnetic field, the identification of wave modes and their speeds, as well as the relationship between them, depend on various plasma parameters. 
The above evolutionarity analysis was performed in the most general form -- the boundaries of the regions in the diagrams are shown schematically, since all quantities along the vertical axis depend on $u_1$ and other quantities ahead of the front.


\bibliographystyle{elsarticle-num} 
\bibliography{amhd,mhd}






\end{document}